\documentclass[12pt,preprint]{aastex}
\input epsf.tex

\begin{document}

\begin{center}
{\bf Galactic Cosmochronometry from Radioactive Elements \\
in the Spectra of Very Old Metal-Poor Stars}

\vspace*{0.3in}
Christopher Sneden \\
Department of Astronomy and McDonald Observatory,
University of Texas, \\
Austin, TX 78712; chris@verdi.as.utexas.edu \\ 

\vspace*{0.2in}
James E. Lawler \\
Department of Physics, University of Wisconsin, \\
Madison, WI 53706; jelawler@facstaff.wisc.edu \\

\vspace*{0.2in}
John J. Cowan \\
Department of Physics and Astronomy, University of Oklahoma, \\
Norman, OK 73019; cowan@mail.nhn.ou.edu \\

\end{center}

\begin{center}
{\bf Abstract}
\end{center}

In a short review of neutron-capture elemental abundances in Galactic
halo stars, emphasis is placed on the use of these elements
to estimate the age of the Galactic halo.
Two prominent characteristics of neutron-capture elements in
halo stars are their large star-to-star scatter in the overall
abundance level with respect to lighter elements, and the dominance 
of $r$-process abundance patterns at lowest stellar metallicities.
The $r$-process abundance signature potentially allows the direct
determination of the age of the earliest Galactic halo nucleosynthesis
events, but further developments in $r$-process theory, high resolution
spectroscopy of very metal-poor stars, and in basic atomic data
are needed to narrow the uncertainties in age estimates.
Attention is brought to the importance of accurate transition
probabilities in neutron-capture element cosmochronometry.
Recent progress in the transition probabilities of rare earth elements
is discussed, along with suggestions for future work on other species.

\section{Introduction}

The very heavy radioactive elements thorium and now uranium have been
detected in metal-poor stars that almost surely were born soon after the 
initial formation of our Galaxy.
The most abundant isotopes of these and almost all other elements with
atomic numbers Z~$>$ 30 are created almost exclusively in neutron
bombardment reactions.
The comparison of Th and U abundances to those of lighter stable
neutron-capture ($n$-capture) elements may in principle be used
to derive age estimates for the Galactic halo.

Much progress has been made in recent years in radioactive element
cosmochronometry, leading to reported metal-poor star age estimates
in the range of 11--16~Gyr.
But uncertainties in these estimates are still large (typically $\pm$3~Gyr).
If this Galactic age technique is to rival other more indirect 
age indicators, further work is needed in many areas: identification of 
suitable candidate metal-poor stars, acquisition of stellar spectra with 
the requisite resolution and signal-to-noise, derivation of accurate
stellar atmosphere models, theoretical prediction of $n$-capture
element production, and determination of reliable atomic transition
parameters for as many $n$-capture species as possible.
In this review we first sketch the general trends in $n$-capture
element abundances in metal-poor stars, and outline the basic techniques
of Th \& U age assessments in stars.
After briefly touching on desired advances in stellar observations,
spectrum analysis, and element predictions, we concentrate on discussing
the application of state-of-the-art atomic transition parameters for
$n$-capture elements to questions of the age and chemical history of
the Galaxy.

\section{Abundance Trends for $n$-Capture Elements in Halo Stars}

Halo stars will be considered here to be those with metallicities 
[Fe/H]~$\leq$ --1.5, where we adopt standard stellar spectroscopic 
definitions that
[A/B]~$\equiv$ log$_{\rm 10}$(N$_{\rm A}$/N$_{\rm B}$)$_{\rm star}$~--
log$_{\rm10}$(N$_{\rm A}$/N$_{\rm B}$)$_{\odot}$, that
log~$\epsilon$(A)~$\equiv$ log$_{\rm 10}$(N$_{\rm A}$/N$_{\rm H}$)~+~12.0,
for elements A and B, and that stellar metallicity may be (arbitrarily) 
assigned to the [Fe/H] value of a star.
The [Fe/H]~$\leq$ --1.5 metallicity limit mostly avoids questions of 
contamination of true Galactic halo stars by members of the thick disk 
populations.
For halo stars, the dominant characteristics of $n$-capture element
distributions are the large star-to-star scatter in their bulk levels 
compared to Fe-peak elements, and their non-solar detailed abundance ratios.

The great variation in overall $n$-capture abundances with respect to lighter
elements is now firmly established [1--4], but with the benefit of
hindsight one can see evidence for this effect in much earlier published
works.
For example, pioneering high resolution studies [5-6] of the bright 
very metal-poor ([Fe/H]~$\simeq$ --2.7) giant HD~122563 found 
[Eu/Fe]~$\simeq$ --0.4, but HD~115444, a star with only slightly smaller
metallicity ([Fe/H]~$\simeq$ --2.9) was found [7] to have 
[Eu/Fe]~$\simeq$ +0.7.
As more very metal-poor stars have been studied, it has become clear that
the range in [$n$-capture/Fe-peak] is at least two orders of magnitude
from star-to-star (e.g., see Figures~14 and 15 in [2]). 
The reality of the scatter can be easily demonstrated by comparisons
of spectra of stars with similar metallicity and atmospheric parameters
but different amounts of $n$-capture elements (e.g. Figure~3 of [4]).
So far no obvious correlation of variable $n$-capture abundance level 
with other stellar characteristic has emerged.
In particular, knowing the [Fe/H] value for a particular halo star 
apparently gives one no immediate insight into its $n$-capture abundances.

Not established yet is the metallicity regime at which the large
variations first appear in [$n$-capture/Fe-peak].
The $n$-capture elements exhibit little star-to-star scatter in high 
metallicity ([Fe/H]~$\geq$ --1.0) disk stars [8], but certainly
significant scatter is evident for stars with ([Fe/H]~$<$ --2.0).
This scatter provides some of the most convincing evidence for the
effects of ``local'' nucleosynthesis events in an apparently poorly
mixed early Galactic halo interstellar medium.

Not only do overall $n$-capture levels vary from star-to-star,
but their abundance distribution in a given star is often very different
than in the solar system.
These elements are almost exclusively produced in neutron bombardment
reactions, but the fusion conditions can vary widely.
In the so-called $s$-process, the neutron flux is weak enough so that
nuclei unstable to $\beta$-decay reactions will have sufficient time
to do so between successive neutron captures.
Synthesis of the $n$-capture nuclei proceeds on well-defined paths along
the valley of $\beta$ stability.
Alternatively in the other extreme labeled the $r$-process, a
huge but short-lived neutron flux temporarily overwhelms $\beta$-decays,
driving nuclei out to the ``neutron drip line''; subsequent $\beta$-decays
after shutoff of the neutron source move the nuclei back toward the
valley of $\beta$ stability.
However, some nuclei are much more easily (sometimes exclusively) 
synthesized in the $s$-process, while others are more abundantly
built by the $r$-process.
This leads to sometimes strikingly different total elemental
abundance distributions from these two types of synthesis events.

\begin{figure}
\epsscale{0.9}
\plotone{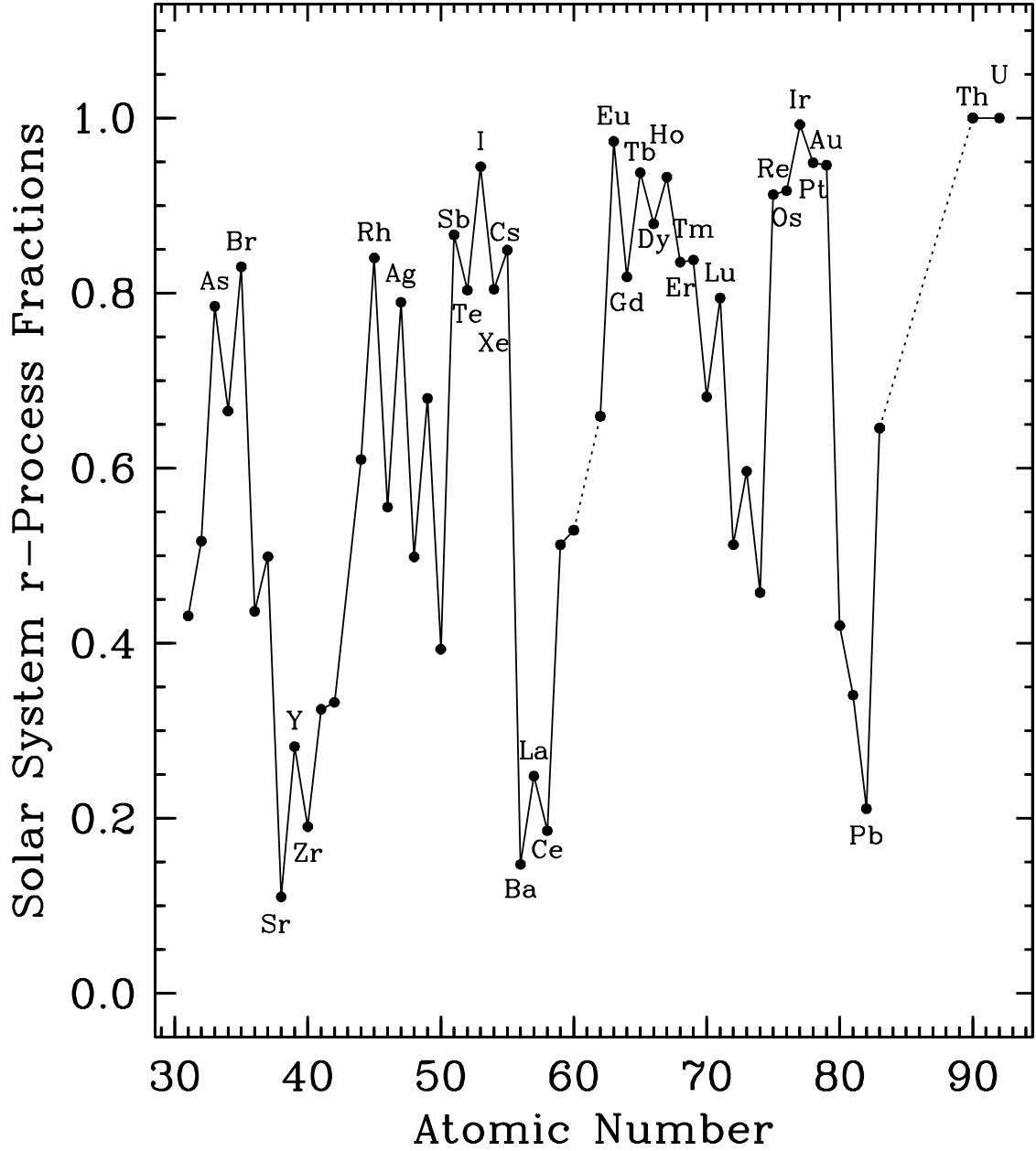}
\caption{
Fractions of solar system $n$-capture abundances due to $r$-process
synthesis.
The data are taken from [4].
Element symbols are written only for those elements with $r$-process
fractional contributions $>$70\% or $<$30\%.
\label{rfrac}}
\end{figure}

In solar-system material, a few easily observed elements appear to
have been created predominantly by one process or the other, while
others seem to be due to a mix of $r$- and $s$-processes.
The breakdown by synthesis process has been studied by several 
groups [9,10,4], and in Figure~\ref{rfrac} we summarize the solar-system 
$r$-process fractional contributions to $n$-capture elements,
adapted from [4].
It is clear from this figure that in the rare-earth element domain,
spectroscopically observable elements such as Ba, La, and Ce were 
predominantly produced in the $s$-process, while Eu, Gd, and Dy were
synthesized by the $r$-process.

Assessment of the relative $r$- and $s$-process contributions to
$n$-capture elements in stars has until recently mostly been a matter
of deriving values of [Ba/Eu], using this ratio to make general
conclusions about the entire $n$-capture element domain.
In doing so, it has been long established [11] that [Ba/Fe] is often 
(usually?) negative at least for stars with [Fe/H]~$<$ --2.
This has led to the conclusion that $r$-process synthesis dominated early
Galactic production of the $n$-capture elements. 
Significant exceptions to this general trend are known to exist
in the discovery of carbon-rich metal-poor stars that have
high ratios of [Ba/Eu], probably indicative of $s$-process activity
in prior evolved intermediate mass stars that was transfered to
the stellar companions observed today (e.g. [12]).
These are interesting stars but are beyond the scope of this review.

Recent detailed spectroscopic studies of some halo stars with sub-solar 
[Ba/Eu] values have established that at least among elements with Z~$>$ 56,
the abundances of all of the observable stable elements are consistent 
with the same sort of $r$-process nucleosynthesis pattern seen in 
solar-system material (e.g.  [13-18]).
This abundance distribution is seen so often that possibly 
nature may know ``only one'' way to both synthesize $n$-capture
elements in conditions of very high neutron flux {\it and} expel 
them into the interstellar medium.
The full $r$-process cannot be reproduced in laboratory conditions,
and is difficult to model theoretically.
Moreover, the astrophysical site of the $r$-process is still not known,
although suggestions are not lacking (Type~II supernovae [19-21];
neutron star mergers [22]; explosive He burning in massive stars [23]).
The repeated pattern of the solar system abundance $r$-process 
distributions in very metal-poor stars provides a severe constraints on 
the entire $r$-process production and distribution scenario.

A dominant $r$-process abundance set of $n$-capture elements observed 
in a star's atmosphere cannot have been created in the interior 
of that star; the explosive nature of the $r$-process precludes it.
Thus, the most $r$-process-rich stars not only have obtained their
$n$-capture elements from a local synthesis event, but that event
in some way (not well established at present) probably was associated
with some aspect of the formation of a neutron star or black hole in a 
supernova explosion.
An interesting observational test might be a radial velocity variability
study of the most $r$-process-rich stars, in order to search for unseen 
compact object companions to these stars.
For example, the extremely $r$-process-rich very metal-poor giant 
CS~22892-052 [15,16] has sinusoidal radial velocity variations with 
an apparent period of 128 days [24].
This is very suggestive of the presence of a companion object, but
unfortunately no mass for the secondary can be estimated from the
data available at present.
Indeed, since the semi-amplitude of the radial velocity variations
is only 1.0~km~s$^{-1}$ [24], the suggested orbital motion of
CS~22892-052 is in need of confirmation.

Further discussion on the mechanics of $r$-process synthesis will not 
be pursued here; see [25] and [26] for further discussion.
Instead the next section discusses cosmochronometry from comparisons
between abundances of stable and radioactively decaying 
$n$-capture elements.

\section{Stellar Ages from Abundances of Thorium and Uranium}

In principle it is straightforward to apply techniques of Th \& U 
cosmochronometry to a star whose $n$-capture elements were created in a 
single prior synthesis event.
First, one must derive accurate abundance of Th, and if possible U also.
Then these abundances need to be compared to those created in the 
original synthesis event, via
$$
N_{\rm X,now}/N_{\rm X,creation} = exp(-t/\tau(X)_{\rm mean}) \\
$$
For example, $\tau$(Th)$_{\rm 1/2}$~= 14.0~Gyr, and after translation to
a mean lifetime,
$$
N_{\rm Th,now}/N_{\rm Th,creation} = exp(-t/20.3Gyr) \\
$$
A similar equation would be written for U, with 
$\tau$(U)$_{\rm 1/2}$ = 4.5~Gyr.

Unfortunately, the geologist's method for establishing $N_{\rm X,creation}$
from measuring the total parent+child elemental abundance cannot be 
employed here. 
The chief decay product of Th and U is Pb, which is very difficult to
observe spectroscopically.
More importantly, Pb does not just grow from Th and U decay, but in fact
is a direct product of the $s$- and $r$-processes.
Therefore unfortunately the initial abundances of Th and U must be 
estimated indirectly.  
Since U has been detected in only one $r$-process-rich star to date (see
below), the usual technique for a star with a measured Th abundance is to 
determine abundances for as many lighter stable $n$-capture elements as 
possible, and then to extrapolate this abundance pattern out to Th.
The first attempt in this manner [27] was done for disk
stars, employing only Nd as a comparison element.
Since Nd is a product of both $s$- and $r$-processes, and since the
Th abundance in a high metallicity disk star is a complex integration
over multiple synthesis events and thus various decay time scales,
most subsequent investigations (e.g. [28])
have concentrated on at least using ``pure'' $r$-process products
such as Eu to compare with Th.

There is some inherent uncertainty in using a stable element nearly 30 
atomic numbers away from Th and U when attempting to probe stellar ages.
Thus, several studies have attempted to push the known $n$-capture
abundance range out to the ``third $n$-capture peak'', Os--Pb.
These are the heaviest stable elements and are much closer in nuclear
mass to Th and U than, for example, the rare-earth elements.
As mentioned in the previous section, some detailed spectroscopic 
investigations of very metal-poor giant stars (CS~22892-052 [16], 
HD~115444 [17], BD+17$^o$3248 [29], and $\simeq$20 other halo giants 
with metallicities [Fe/H]~$\leq$ --1.6 [18])
have yielded observed abundance distributions that are very good 
matches to the solar system $r$-process-only abundances.
This suggests that one may extrapolate the solar system curve out to
the actinide element domain to establish the zero-decay-age Th 
abundance for these stars.
Doing this yields implied ages in the range 11--16~Gyr.
The $\simeq$5~Gyr range is large enough to be consistent with more 
indirect methods for Galactic age estimates, such as globular cluster 
color-magnitude diagram modeling.

There are some important uncertainties to keep in mind in this
kind of exercise.
First, the age computation involves an exponential, so that any small
error in the observed Th abundance translates to a large age error.
Second, it is not easy to conclusively establish that the extrapolation
of the solar system abundance distribution to Th is correct for
the $r$-process events that occurred in the early Galaxy.
Indeed some have argued [30] that the production of Th is sensitive 
to many details of the synthesis event, and is difficult
to predict even with knowledge of the abundances of many lighter
elements.
On the other hand, investigation of nuclear models [25] suggests
that the production of Th is not alterable at will,
and in fact is reasonably the same for realistic simulations of 
$r$-process synthesis events.

A major new observational development has been the detection [31] of 
not only Th but also U in the very metal-poor giant CS~31082-001.
The Th/U production ratio probably can be estimated with much greater 
reliability than can Th/Eu or similar ratios of Th with respect to lighter 
stable elements (see [29] for further discussion of this point).
The observed Th/U ratio in CS~31082-001 (which should be sensitive 
function of age, given the very different decay rates of Th and U), 
is consistent with an age of 12-13~Gyr for this star [31].
This agreement with the Th-based ages for other metal-poor stars
is encouraging, but detection of U in more stars would obviously be
welcome.

Since the present review emphasizes the observational aspects of radioactive
element cosmochronometry, detailed discussion about theoretical
predictions of Th and U abundances will be deferred to other papers
(e.g. [32]).
Given the difficulty in making $r$-process predictions from laboratory
data, and the enormous range in $n$-capture abundance levels in 
metal-poor stars, radioactive element cosmochronometry probably will be 
shown to yield reliable ages of Galactic halo stars only when the 
distribution of the entire range of $n$-capture elements is understood 
in a large number of stars over the entire halo metallicity range.

A future goal of $r$-process observational studies will be to discover 
whether or not significant exceptions to the solar-system $r$-process 
abundance pattern exists in halo stars.
The exceptions, or possibly lack thereof, will empirically establish 
the limits that nature imposes on $r$-process production in the Galaxy.
If the solar-system pattern pattern extends over all stable heavy $n$-capture
elements in $r$-process-rich stars, then abundances of unstable Th and U
may be compared with confidence to $any$ of the stable elements with 
Z~$>$ 56 in cosmochronometry studies.
If the pattern is often broken, only comparisons of Th and U to the 
closest neighboring stable elements in the periodic table can yield 
age estimates with confidence.
In the next sections we will discuss two aspects of this problem: 
obtaining the stellar spectra, and improving the laboratory data of 
important $n$-capture species.

\section{Candidate Stars for Cosmochronometry}

Since the goal is to use stellar abundances of thorium and uranium to 
derive radioactive decay-based ages for the Galaxy, samples of stars are
needed with the following characteristics: (a) birth epochs as near as
possible to the formation of the Galaxy; (b) evolved evolutionary states;
(c) relatively high abundance ratios of $n$-capture elements to those
of the Fe-peak; and (d) apparent brightness large enough to permit 
the requisite spectra to be acquired.

Galactic halo stars provide the obvious target sample, but there
may exist a significant age range in the general halo field population.
Color-magnitude diagram analysis can provide age estimates for stars that 
are members of clusters, but this is not possible for field stars.
Metallicity and kinematics are observationally the best stellar
age surrogates, and so major surveys have concentrated in these areas.
Metallicities for nearby halo stars have been determined in a
large survey [33] of high proper motion stars. 
Parallaxes exist for many of these stars, yielding the definite
advantage of the ability to compare abundance characteristics with
kinematics.
However, these stars are most often relatively unevolved main sequence 
dwarfs, turnoff stars, and subgiants.
As such, they have relatively high gravities, and thus have narrow,
high density atmospheres that produce weak-lined spectra that favor
transitions of neutral atoms over the ionized species that account
for the majority of $n$-capture element lines in the visible
spectral domain.

Giants should be better candidates for detections of Th and U in 
halo stars, because they have deep, low density atmospheres that produce 
much stronger ionized line spectra than do the higher gravity stars.
A Galactic pole region objective prism survey ([34] and references 
therein) has provided an appropriate (and very large) sample of 
low metallicity giants far out into the Galactic halo.
This stellar database is now being explored at medium and high spectral
resolution to elucidate the abundance trends over a range of metallicities.
Most attention to date has centered so far on stars with [Fe/H]~$<$ --2.5
in this sample, since higher metallicity domains have been
studied extensively with target lists of more nearby (and brighter) 
field halo giants (e.g. [35])

\begin{figure}
\epsscale{0.9}
\plotone{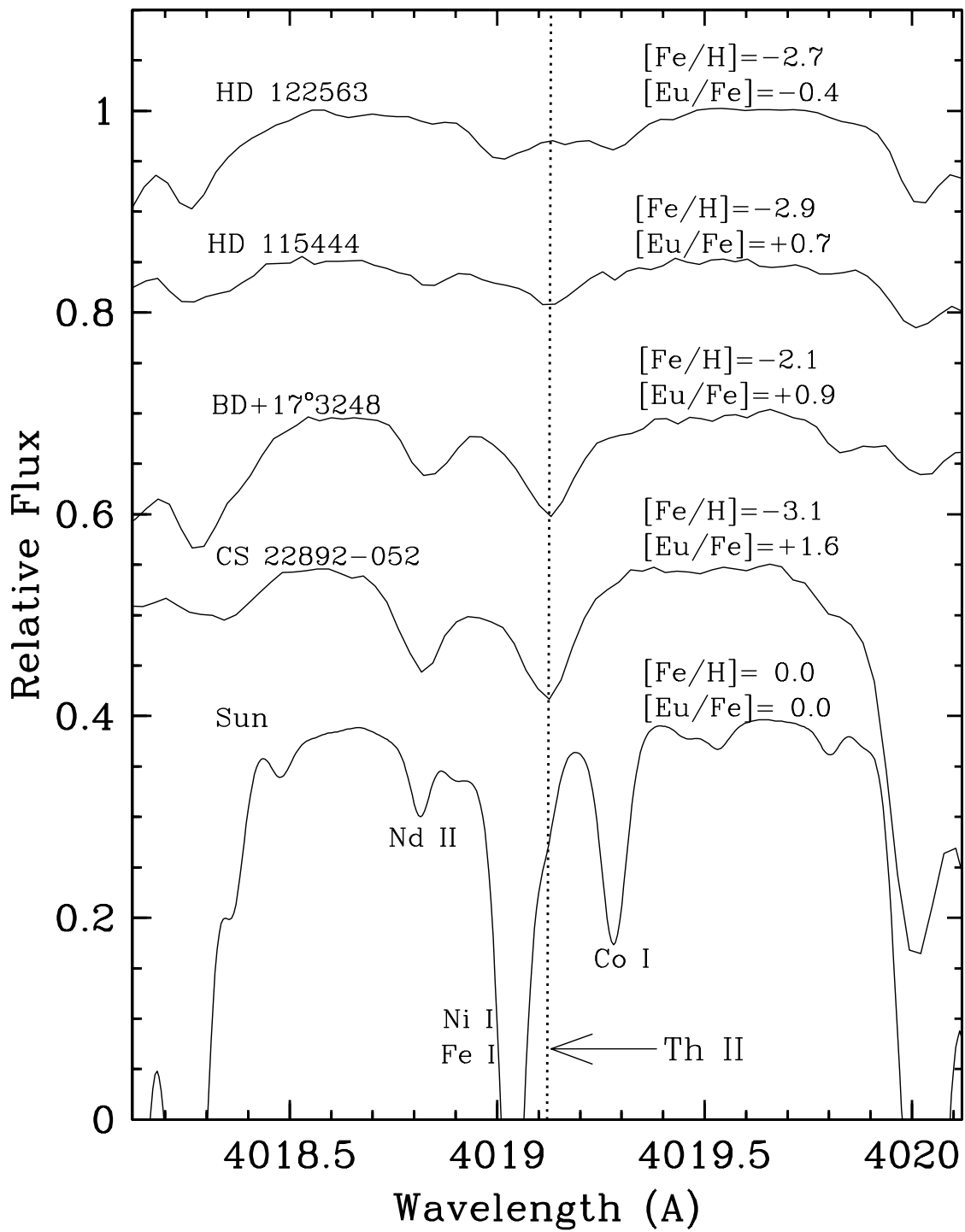}
\caption{
Spectra of the Th~II 4019.12~\AA\ line in the Sun and four very metal-poor
stars.
The spectrum sources are: HD~122563 and HD~115444, [17]; 
BD+17$^o$3248, [29]; CS~22892-052, [16]; and the Sun, [49].
The [Fe/H] and [Eu/Fe] values of the four stars are also taken from 
these papers.
The wavelength of the Th~II line is marked with a dotted line, and a few
other atomic features are also labeled.
\label{spthorium}}
\end{figure}

Of the Galactic pole giant star sample, only low metallicity stars 
with ``super-solar" $n$-capture elemental abundances are suitable 
for Th and U age estimates.
The reason is simply that only those stars with 
[$n$-capture/Fe-peak]~$\gg$~0 AND [$n$-capture/C]~$\gg$~0 can produce 
Th~II and U~II features that are significantly stronger
than the large number of contaminating lines that surround them.
We illustrate this effect in Figure~\ref{spthorium}, which shows the spectrum 
of the 4019.12~\AA\ Th~II line in the Sun and four metal-poor giants.
The figure notes not only the stellar [Fe/H] metallicity but also
the [Eu/Fe] value, standing as a simple measure of the overall
[$n$-capture/Fe-peak] level.
The 4019~\AA\ line is by far the strongest Th~II line in cool stars, but 
unfortunately it is part of a complex atomic and molecular feature blend, 
which probably includes lines of V~I, Mn~I, Fe~I, Fe~II, Co~I, and Ni~I 
[28,36], perhaps Tb~II and/or Ce~II [15], and $^{13}$CH [37].
Absorption from the Th~II line is visible in the solar spectrum displayed 
at the bottom of this figure, but it comprises only a small fraction of 
the total feature at this wavelength. 
Very accurate Th abundances may not easily be determined from the 
solar photospheric spectrum.

The Th~II line cannot detected at all in the halo giant HD~122563 
(Figure~\ref{spthorium}, top spectrum), which has been the subject 
of numerous abundance analyses.
This star has an underabundance not only of Eu as indicated in the
figure, but of all of the heavier $n$-capture elements ([17], and 
references therein).
That paper showed through comparisons of synthetic and observed Th~II spectra 
that the entire 4018.9--4019.3~\AA\ absorption can be accounted for 
without invoking the presence of the Th~II line.
The $n$-capture-rich stars shown as the middle three spectra of
Figure~\ref{spthorium} (HD~115444 [17]; BD+17$^o$3248 [18,29]; 
CS~22892-052, [15,16]) have clearly detected Th~II lines.
However, even in these cases the Th~II absorption is not isolated, and
significant contamination still exists.  
Particular attention needs to be paid to the $^{13}$CH lines in C-rich
stars such as CS~22892-052; in some cases the $^{13}$CH absorption can 
overwhelm the rest of the components of the 4019~\AA\ blend [37].

Finally, the spectra displayed in Figure~\ref{spthorium} demonstrate that 
in almost all cases the Th~II line will attain a central depth of $\leq$10\%.
This weakness combined with the complexity of the contaminant features
mandates the acquisition of high resolution (R~$\lambda/\Delta\lambda$~$>$
40,000), high signal-to-noise (S/N~$>$ 100) if reliable Th abundances
are to be obtained from the 4019~\AA\ line.
Other Th~II lines, e.g. 4086.6~\AA, are detectable if the Th abundance
is very large, but they are all weaker and have their own line contamination
problems.

The observational situation for U is even more difficult.  
The only reasonably strong U~II line in most cool stellar spectra, 3859.6~\AA,
lies in the middle of the strong CN blue system $\Delta$v~= 0 sequence.
Therefore the U~II line will be totally obliterated in most spectra of
cool stars.
The presence of this line in CS~31082-001 (Figure~1 of [31])
is a happy but probably rare conjunction of very low metallicity 
([Fe/H]~= --3.1), effectively absent CN bands near 3860~\AA, and
extreme overabundances of all elements with Z~$\geq$ 56 (e.g., 
[Os,Ir/Fe]~$\sim$ +2.0).

\section{Atomic Transition Data}

Until very recently, $n$-capture abundance analyses had to use very scattered
laboratory transition probabilities for many species, and several important
transitions had no reliable $gf$ values at all.
Moreover, the necessary hyperfine and isotopic structure information 
needed to derive reliable abundances from strong (saturated) lines was
sparse.
Happily, several atomic physics groups are now publishing extensive data
for especially the rare earths, and in the process bringing the transition
parameters to the level required by the new high resolution stellar spectra.

In this section we will discuss some of these new data, and suggest some
needs for future work in this area.
Pending publication of the full abundance results for CS~31082-001 [31],
the $r$-process-rich ultra-metal-poor giant CS~22892-052 is the halo star 
with the most detections of $n$-capture elements.  
Nearly all other metal-poor stars are likely to yield fewer detections.
and the combination of element detections in CS~22892-052 [16]
and HD~115444 ([17]) should cover nearly all $n$-capture 
elements that are likely to be analyzed in metal-poor star spectra.
We concentrate on those species discovered in those two stars.

The discussion of new atomic data will emphasize the new results of the 
Wisconsin group, but excellent data are being published by other
atomic physics teams as well.
A good survey of literature data for lines arising from lowest energy
states of $n$-capture species (often those lines that are most easily 
observed) is given in [38].
The reader is urged to consult that discussion for many elements.
Some further comments, and attempts to reconcile some conflicting
literature $gf$ data are in [15].
Consulting these and other sources can give an idea of which species
have well-established $gf$ scales and which need work.

\begin{figure}
\epsscale{0.9}
\plotone{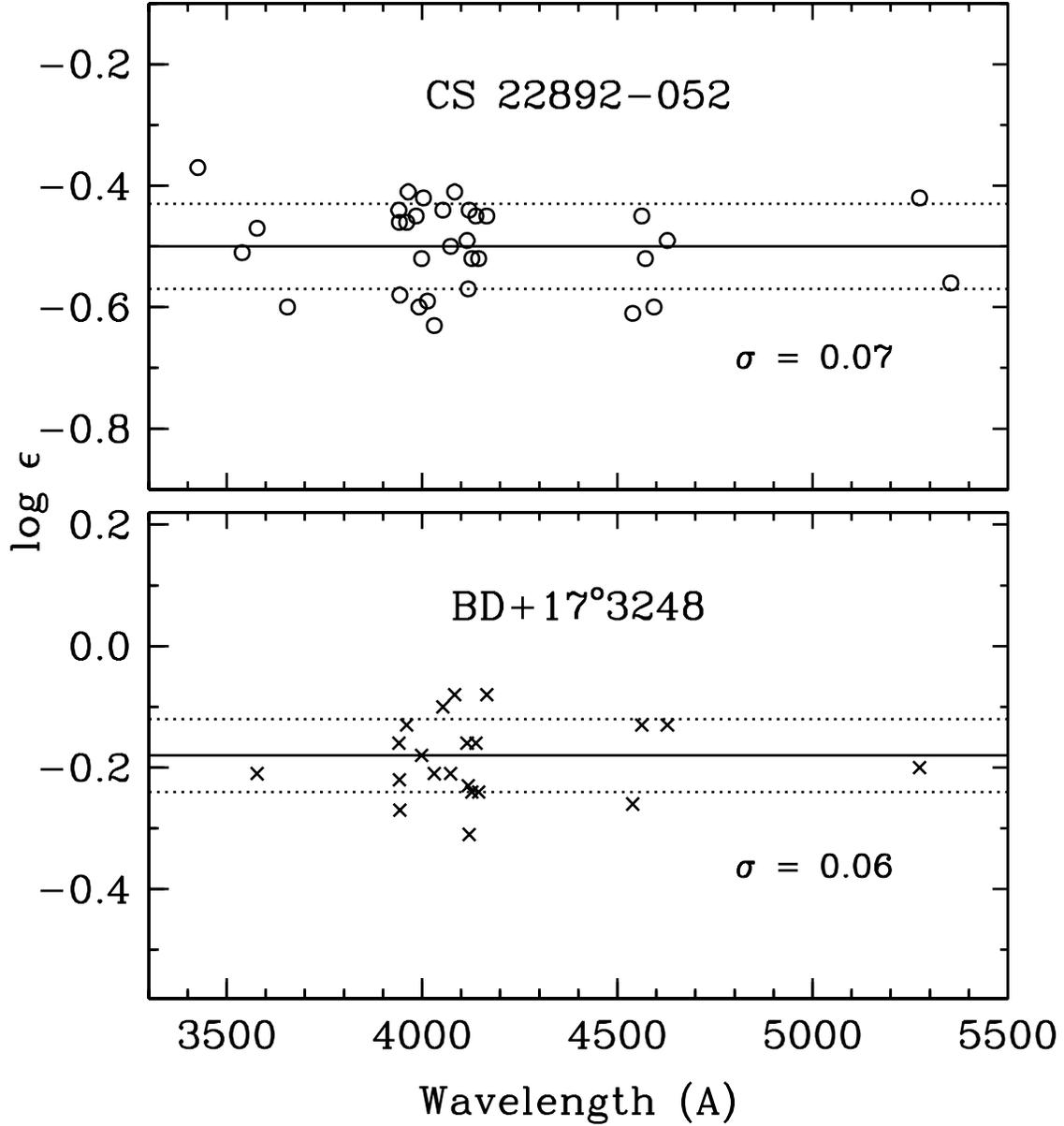}
\caption{
Individual Ce II line abundances for the very metal-poor, $r$-process-rich 
stars CS~22892-052 (top panel) and BD+17$^o$3248 (bottom panel).
The $gf$ values for the transitions are taken from [39], and the 
stellar spectra employed in the analysis were those
cited in Figure~\ref{spthorium}.
For each star the mean Ce abundance is shown as a solid horizontal line.
The sample standard deviation value $\sigma$ is both written in
the figure panel and shown as dotted lines displaced from the mean abundance
line.
\label{cerium}}
\end{figure}

As an opening example of the recent improvements in data for
ionized rare earths, consider the case of Ce~II.
The combination of experimental lifetime measurements and theoretical 
calculations have produced [39] a very extensive set of $gf$ values for Ce~II.
This species is potentially interesting because in solar-system material
Ce, like Ba, is predominantly a product of the $s$-process.
But Ba~II presents only 4--5 very strong transitions and a couple of
very weak ones for abundance analyses, whereas Ce~II lines of moderate
strength can be found throughout the entire near UV to visible spectrum.
We ``tested'' the reliability of the new $gf$'s by applying them to the
spectra of CS~22892-052 [16] and BD+17$^o$3248 [29].
Equivalent width measurements were made only for easily identified lines
in these spectra, and abundances were derived in the manner described
in the papers on these stars.  
The abundances derived for each Ce~II line in the two stars are shown 
in Figure~\ref{cerium}, plotted as functions of wavelength (plots with
respect to line excitation potential, $gf$, or equivalent width revealed
no trends in abundances with respect to these quantities).
The key point shown in this figure is the very small line-to-line
scatter ($\sigma$~= 0.06--0.07) in the abundances.
For comparison, the study of CS~22892-052 [15] with lower quality spectra 
and transition probabilities only employed nine Ce~II lines, and the 
resulting line-to-line abundance scatter, $\sigma$~= 0.16, was more 
than double the present values.

The Ce~II data demonstrate the necessity of having transition probability
data for many lines of a species that at least have the internal 
consistency to yield stellar abundances with small line-to-line scatter.
Then for those stars with many measurable lines of a species, the average 
elemental abundance values can be meaningfully employed to address larger 
issues, in particular to compare the observed abundances with theoretical 
nucleosynthesis predictions.
This can be accomplished for stars with large [$n$-capture/Fe] ratios, 
which present rich spectra of the $n$-capture elements.
Just as importantly, the existence of internally consistent $gf$ sets
allow the derivation of accurate abundances in stars with lower
$n$-capture element levels; such stars often may exhibit only a few
detectable transitions of a species.

\begin{figure}
\epsscale{0.9}
\plotone{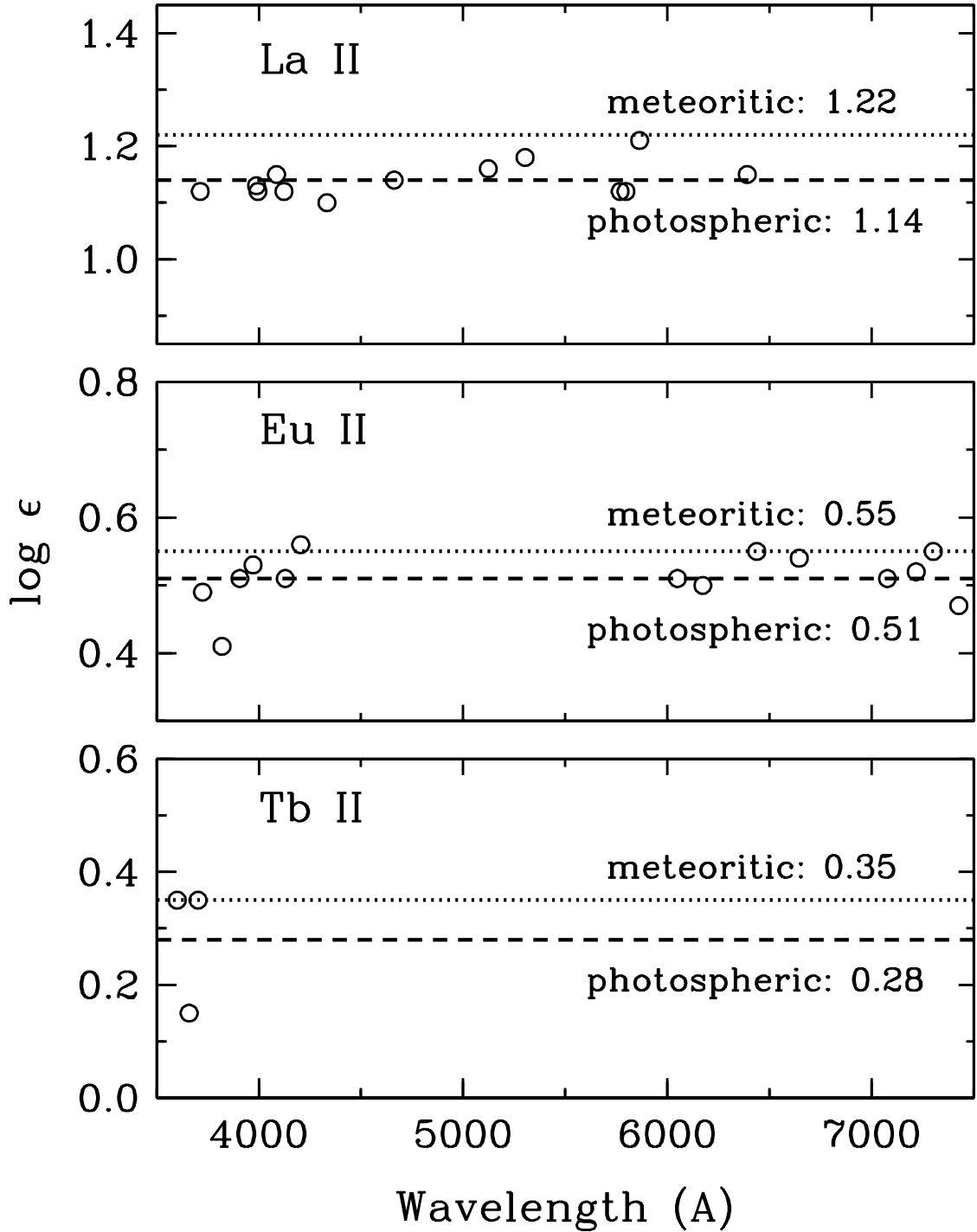}
\caption{
Solar photospheric abundances from individual lines of La~II, Eu~II, 
and Tb~II, using the new atomic data described in the text.
The photospheric abundances written in the figure panels are
also indicated by dashed lines.
For comparison, the meteoritic values recommended by [43] are 
shown as dotted lines.
\label{sollines2}}
\end{figure}

The Wisconsin atomic physics group has been gathering lifetime data and
very high resolution (Fourier transform) spectra for several rare
earth elements, combining these data to produce $gf$ values and 
hyperfine/isotopic parameters for many lines of neutral atoms
and their first ions.
These data are then applied to the solar spectrum to try to improve the 
(sometimes poorly determined) photospheric abundances of these elements. 
The most recent efforts have yielded extensive new atomic data
for La~II [40], Eu~II [41], and Tb~II [42].
In Figure~\ref{sollines2} we summarize the solar abundance results using
these data; see the papers for descriptions of the analysis techniques, which
typically involved matching synthetic and observed solar spectra.
The line-to-line abundance scatters are satisfactorily small 
($\sigma$~$\simeq$ 0.04) for La and Eu, but not so for Tb.
The problem for Tb lies not in the atomic data but in the extreme
weakness of all Tb~II transitions in the solar spectrum.
There appear to be no useful lines of Tb~II beyond 4000~\AA, and
only three very weak, partially blended lines in the crowded near-UV
solar spectrum.
Even so, this new photospheric Tb abundance is in much closer
accord with the meteoritic value [43] than previously estimated.
For all three of these rare earth elements, the consistently small 
negative offset, averaging -0.06~dex, of the solar photospheric abundances 
with respect to meteoritic abundances is not well understood and 
deserves further study.

For cosmochronometry studies, one is of course concerned with the state
of Th~II and U~II transition data, which fortunately appear to be 
reasonably well established for critical features of these species.  
The Th~II 4019~\AA\ line was studied in [44], and the 
U~II 3859~\AA\ transition probability has been reconsidered in [45].
More extensive lab studies of other Th~II transitions would be useful,
as they become detectable in the most $r$-process-rich stars [31].

\begin{figure}
\epsscale{0.9}
\plotone{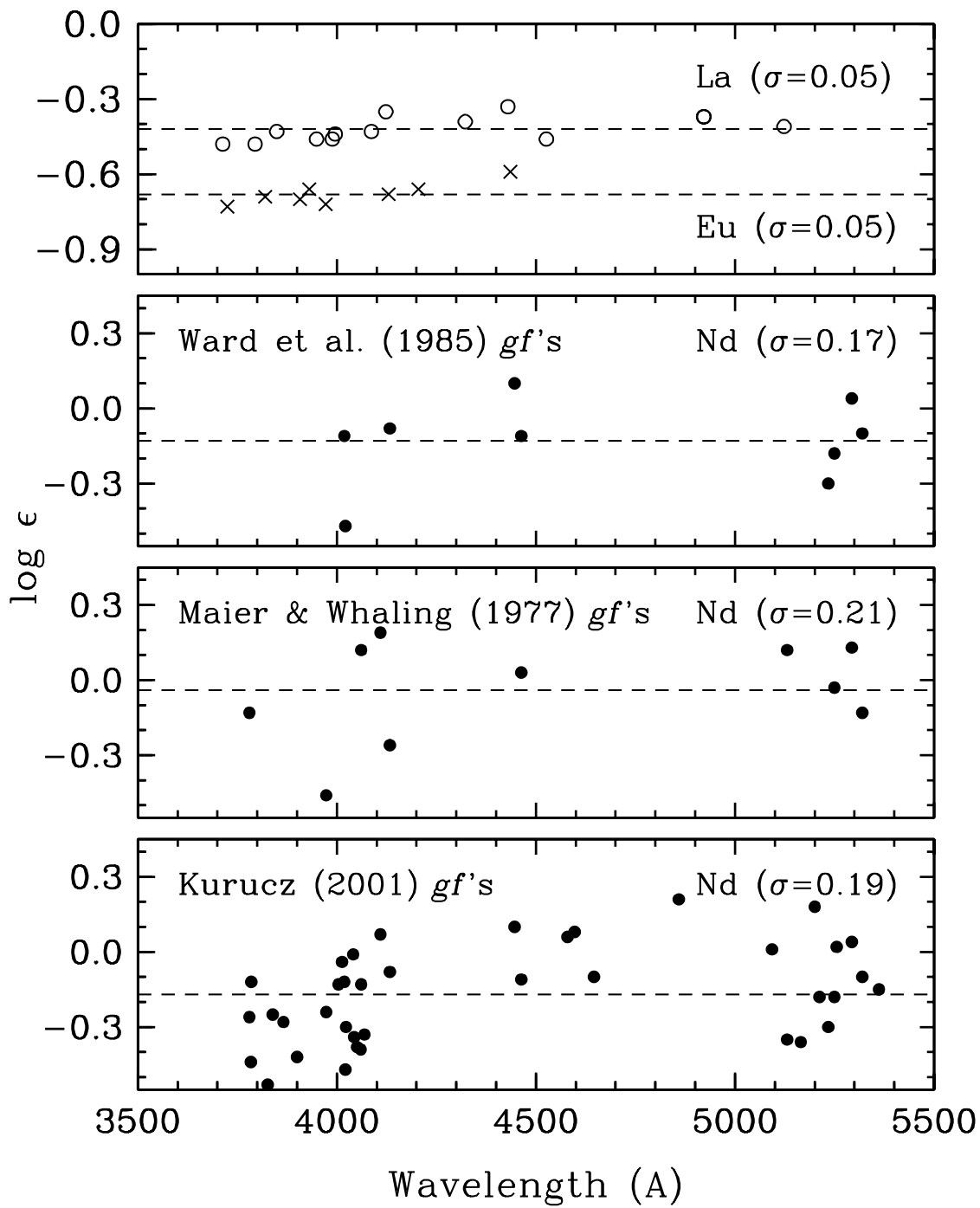}
\caption{
Abundances of La~II, Eu~II, and Nd~II in BD+17$^o$3248.
In the top panel, individual line abundances for La and Eu are shown 
as open circles and $\times$ symbols, the mean abundances are shown 
as dashed lines, and the $\sigma$ values are noted.
These abundances have been generated using the atomic data of
[40,41].
The other three panels show the same kinds of data for Nd~II lines
in BD+17$^o$3248, adopting three different sets of transition probabilities.
\label{bdscat2}}
\end{figure}

In spite of the recent progress, some species relevant to cosmochronometry
still need attention.
Several neutral atoms of the so-called ``third $n$-capture peak''
have only a few trustworthy $gf$'s, and frustratingly those transitions
tend to occur in the UV ($\lambda$~$<$ 3200~\AA) spectral domain,
difficult to observe and to model.
A couple ionized species of rare earths have poorly understood 
transition probabilities, and as an example consider data for 
Nd~II in BD+17$^o$3248.
In Figure~\ref{bdscat2} the well-determined abundances of La and Eu
(computed with the new atomic data discussed above) are displayed in the 
top panel, and results of abundance computations for Nd with three
different sets of published $gf$ values are shown in the other
panels.
The $gf$ values of [46] and [4] are derived from experimental data, while
those of [48] adoptions or corrections of the values in the literature.
It is easily seen that none of these atomic data sets produce satisfactory
abundance results for Nd in BD+17$^o$3248, and the large $\sigma$ values 
appear to be unrelated to the number of lines used in the analyses.
An attempt to reconcile the $gf$-scales of [46] and [47]
in [15] produced equally unsuitable results.
This species clearly deserves a new laboratory analysis.

Aside from the problem of Nd~II, we recommend renewed attention
to neutral atomic lines of the Os--Pb element group and elements
in the range 41~$\geq$ Z~$\geq$ 48; both of these sets of elements are 
just now beginning to be spectroscopically accessible in metal-poor stars,
and often the transition data are older and less extensive than the
data for the rare earths.

\section{Summary}

Establishing an age for the Galactic halo from observations of Th and U 
transitions in metal-poor stars demands progress in a number of areas.
We have emphasized here that obtaining abundances for the radioactive 
elements for this task is only one part of the problem.
Greater understanding of the production of the whole range of $n$-capture 
elements is essential, and this task cannot be accomplished without
the best possible astronomical spectra of $n$-capture-rich halo
stars and continuing progress on laboratory data.
Nucleosynthesis theory can be confronted only at the level of accuracy
that can be established by abundance analyses.
Improving atomic data and stellar spectra go jointly forward in
this endeavor.

\acknowledgements

We are pleased to thank the US National Science Foundation for its
support of this work, most recently through grants AST-9987162 to C.S.,
AST-9819400 to J.E.L., and AST-9986974 to J.C.C.

\clearpage

\begin{center}
{\bf REFERENCES}
\end{center}

\noindent 1. Gilroy, K. K., Sneden, C., Pilachowski, C. A., 
\& Cowan, J. J., Astrophys. J., {\bf 327}, 29 (1988).

\noindent 2. McWilliam, A., Preston, G. W., Sneden, C., \& Searle, L., 
Astron. J., {\bf 109}, 2757 (1995).

\noindent 3. Ryan, S. G., Norris, J. E., \& Beers, T. C., Astrophys. J., 
{\bf 471}, 254 (1996).

\noindent 4. Burris, D. L., Pilachowski, C. A., Armandroff, T. A., 
Sneden, C., Cowan, J. J., \& Roe, H., Astrophys. J., {\bf 544}, 302 (2000).

\noindent 5. Wallerstein, G., Greenstein, J. L., Parker, R., Helfer, H. L., 
\& Aller, L. H., Astrophys. J., {\bf 137}, 280 (1963).

\noindent 6. Pagel, B. E. J., Roy. Obs. Bull., {\bf 104}, 127 (1965).

\noindent 7. Griffin, R., Griffin, R., Gustafsson, B., \& Vieira, T., 
Mon. Not. Roy.  Astron. Soc., {\bf 198}, 637 (1982).

\noindent 8. Woolf, V. M., Tomkin, J., \& Lambert, D. L., Astrophys. J., 
{\bf 453}, 660 (1995).

\noindent 9. Cameron, A. G. W., Astrophys. Space Sci. {\bf 82}, 123 (1982).

\noindent 10. K{\"a}ppeler, F., Beer, H., \& Wisshak, K., Rep. Prog. Phys., 
{\bf 52}, 945 (1989).

\noindent 11. Spite, M., \& Spite, F., Astron. Astrophys., 
{\bf 67}, 23 (1978).

\noindent 12. Norris, J. E., Ryan, S. G., \& Beers, T. C., Astrophys. J., 
{\bf 488}, 350 (1997).

\noindent 13. Sneden, C., \& Parthasarathy, M., Astrophys. J., 
{\bf 267}, 757 (1983).

\noindent 14. Sneden, C., \& Pilachowski, C. A., Astrophys. J., 
{\bf 288}, L55 (1985).

\noindent 15. Sneden, C.,  McWilliam, A., Preston, G. W., Cowan, J. J., 
Burris, D. L., \& Armosky, B. J., Astrophys. J., {\bf 467}, 819 (1996).

\noindent 16. Sneden, C., Cowan, J. J., Ivans, I. I., Fuller, G. M., 
Burles, S., Fuller, G., Beers, T. C., \& Lawler, J. E., Astrophys. J., 
{\bf 533}, L139 (2000).

\noindent 17. Westin, J., Sneden, C., Gustafsson, B., \& Cowan, J. J., 
Astrophys. J., {\bf 530}, 783 (2000).

\noindent 18. Johnson, J. R., \& Bolte, M., Astrophys. J., 
{\bf 554}, 888 (2001).

\noindent 19. Woosley, S. E., Wilson, J. R., Mathews, G. J., 
Hoffman, R. D., \& Meyer, B. S., Astrophys. J., {\bf 433}, 229 (1994).

\noindent 20. Takahashi, K., Witti, J., \& Janka, H.-T., 
Astron. Astrophys., {\bf 286}, 857 (1994).

\noindent 21. Mathews, G. J., Bazan, G., \& Cowan, J. J.. 
Astrophys. J., {\bf 391}, 719 (1992).

\noindent 22. Rosswog, S., Liebendörfer, M., Thielemann, F.-K., 
Davies, M., Benz, W., \& Piran, T., Astron. Astrophys., 
{\bf 341}, 499 (1999).

\noindent 23. Thielemann, F.-K., Arnould, M., \& Hillebrandt, W.,
Astron. Astrophys., {\bf 74}, 175 (1979).

\noindent 24. Preston, G. W., \& Sneden, C., Astron. J., in press (2001).

\noindent 25. Cowan, J. J., Pfeiffer, B., Kratz, K.-L., Thielemann, F.-K., 
Sneden, C., Burles, S., Tytler, D., \& Beers, T. C., Astrophys. J., 
{\bf 521}, 194 (1999).

\noindent 26. Freiburghaus, C., Rembges, J.-F., Rauscher, T., Kolbe, E., 
Thielemann, F.-K., Kratz, K.-L., Pfeiffer, B., \& Cowan, J. J., 
Astrophys. J., {\bf 516}, 381 (1999).

\noindent 27. Butcher, H. R., Nature, {\bf 328}, 127 (1987).

\noindent 28. Fran\c{c}ois, P., Spite, M., \& Spite, F., 
Astron. Astrophys., {\bf 274}, 821. (1993).

\noindent 29. Cowan, J. J., Sneden, C., Burles, S., Ivans, I. I., 
Truran, J. W., Beers, T. C., Primas, F., Fuller, G. M., \& Kratz, K.-L., 
Astrophys. J., submitted (2001).

\noindent 30. Goriely, S. \& Clerbaux, B.  Astron. Astrophys., 
{\bf 346}, 798 (1999).

\noindent 31. Cayrel, R., Hill, V., Beers, T. C., Barbuy, B., Spite, M., 
Spite, F., Plez, B., Andersen, J., Bonifacio, P., Fran\c{c}ois, P., 
Molaro, P., Nordstrom, B., \& Primas, F., Nature, {\bf 409}, 691 (2001).

\noindent 32. Burles, S., Truran, J. W., Cowan, J. J., Sneden, C., 
Kratz, K.-L, \& Pfeiffer, B., preprint (2001).

\noindent 33. Carney, B. W., Latham, D. W., Laird, J. B., 
\& Aguilar, L. A., Astron. J., {\bf 112}, 668 (1994).

\noindent 34. Beers, T. C., Preston, G. W., \& Shectman, S. A., Astron. J., 
{\bf 103}, 1987 (1992).

\noindent 35. Bond, H. E., Astrophys. J. Suppl., {\bf 44}, 517 (1980).

\noindent 36. Morell, O., K{\"a}llander, D., \& Butcher, H.~R., 
Astron. Astrophys., {\bf 259}, 543 (1992).

\noindent 37. Norris, J. E., Ryan, S. G., \& Beers, T. C., Astrophys. J., 
{\bf 489}, L169 (1997).

\noindent 38. Morton, D. C., Astrophys. J. Supp., (2000).

\noindent 39. Palmeri, P., Quinet, P., Wyart, J.-F., \& Bi\'emont, E., 
Phys. Scr., {\bf 61}, 323 (2000).

\noindent 40. Lawler, J. E., Bonvallet, G., \& Sneden, C., Astrophys. J., 
{\bf 556}, 452 (2001).

\noindent 41. Lawler, J. E., Wickliffe, M. E., Den Hartog, E. A. 
\& Sneden, C., Astrophys. J., in press (2001).

\noindent 42. Lawler, J. E., Wickliffe, M. E., Cowley, C. R. \& Sneden, C., 
Astrophys. J., in press (2001b).

\noindent 43. Grevesse, N., \& Sauval, A. J., Sp. Sci. Rev., 
{\bf 85}, 161 (1998).

\noindent 44. Lawler, J.~E., Whaling, W., \& Grevesse, N., Nature, 
{\bf 346}, 635 (1990).  

\noindent 45. Lundberg, H., Johansson, S., Nilsson, H., \& Zhang, Z., 
Astron. Astrophys. J., {\bf 372}, 50 (2001).

\noindent 46. Maier, R.~S. \& Whaling, W., J. Quant. Spec. Rad. Trans., 
{\bf 18}, 501 (1977).

\noindent 47. Ward, L., Vogel, O., Arnesan, A., Hallin, R., 
\& W{\"a}nnstr{\"o}m, A., Phys. Scr., {\bf 31}, 161 (1985).
 
\noindent 48. Kurucz, R. L., http://cfaku5.harvard.edu/ (2001).

\noindent 49. Kurucz, R. L., Furenlid, I., Brault, J., \& Testerman, L., 
``Solar Flux Atlas from 296 to 1300 nm'' (Harvard University Press,
Cambridge, Massachusetts 1984).

\clearpage

\clearpage

\clearpage

\clearpage

\clearpage

\end{document}